 \documentclass[12pt]{iopart}
 
\usepackage{iopams}
\usepackage{amssymb}
\usepackage{amsgen}

\usepackage[english]{babel} \usepackage{latexsym}
\usepackage{graphicx} \usepackage{subfigure} \usepackage{epsfig}
\usepackage{psfrag}

\begin{document}

\title[Phase transition from straight into twisted vortex-lines
in dipolar BEC]{Phase transition from straight into twisted vortex-lines
in dipolar Bose-Einstein condensates}

\author{M. Klawunn and L. Santos}

\address{Institut f\"ur Theoretische Physik, Leibniz Universit\"at
Hannover, Appelstr. 2, D-30167, Hannover, Germany}

\begin{abstract}
The non-local non-linearity introduced by 
the dipole-dipole interaction plays a crucial role in the 
physics of dipolar Bose-Einstein condensates. In particular, 
it may distort significantly the stability of straight vortex 
lines due to the rotonization of the Kelvin-wave spectrum. 
In this paper we analyze this instability showing that it 
leads to a second-order-like phase transition from a straight vortex-line into novel 
helical or snake-like configurations, depending on the dipole orientation. 
\end{abstract}
\submitto{\NJP}



\section{Introduction}
\label{sec:Intro}


The physics of ultracold atomic and molecular gases is crucially 
determined by the interparticle interactions. Typical experiments 
in quantum gases have studied up to now particles which interact 
dominantly via short-range isotropic potentials. At the very low energies
involved in these experiments, such interactions are characterized 
by a single parameter, namely the $s$-wave scattering length. 
However, a new generation of recent experiments is opening a fascinating 
new research area, namely that of dipolar gases, for which the dipole-dipole 
interaction (DDI) plays a significant or even dominant role. 
These experiments include on one side those dealing with the DDI effects due to  
magnetic dipoles in degenerate atomic gases, as it is the case 
of recent exciting experiments in Chromium 
Bose-Einstein condensates (BECs) \cite{Chromium} 
and Rubidium spinor BECs \cite{Vengalattore}. 
On the other side, recent experiments on the creation of heteronuclear
molecules in the lowest vibrational states \cite{Molecules}, although not yet 
brought to quantum degeneracy, open fascinating perspectives for the generation
of polar molecules with very large electric dipole moments. Last but not
least, the DDI in Rydberg atoms is extremely large \cite{Rydberg} and may
allow for e.g. the construction of fast quantum gates \cite{Jaksch}.


Contrary to the isotropic van-der-Waals-like short-range interactions, 
the DDI is long-range and anisotropic 
(partially attractive), and leads to fundamentally new physics in 
ultra cold gases, modifying e.g. the stability and excitations of BECs
\cite{Stability,Roton}, the properties of Fermi gases \cite{Fermions}, and  
the physics of strongly-correlated gases \cite{DipLat-FQHE}.  
Time-of-flight experiments in Chromium condensates allowed for the first
observation of DDI effects in quantum degenerate gases \cite{Expansion}, 
which have been remarkably enhanced recently by means of 
Feshbach resonances \cite{Pfau_new}. In addition, the DDI has been recently
shown to play an important role in the physics of spinor Rubidium BEC 
\cite{Vengalattore}, and to lead to an observable damping of 
Bloch oscillations of Potassium atoms in optical lattices \cite{Fattori}.  


The long-range character of the DDI leads to nonlocal nonlinearity in dipolar
BECs, similar as that encountered in e.g. 
plasmas \cite{Plasma} and nematic liquid crystals \cite{Nematics}. 
This nonlocality leads to novel nonlinear physics in dipolar BECs,
including the possibility of obtaining stable 2D bright solitons 
\cite{BrightSolitons} and stable 3D dark solitons \cite{DarkSolitons}. 
In addition the partially attractive character of the nonlocal nonlinearity 
due to the anisotropy of the DDI has remarkable consequences 
for the stability of dipolar BECs, which may become unstable against collapse 
in 3D traps, as recently shown in experiments with Chromium BECs \cite{PfauCollapse}.  
On the contrary 2D traps may allow for an instability without collapse,
characterized by the formation of a gas of inelastic 2D solitons \cite{Phonon}.


Vortices and vortex-lines constitute a prominent topic among
the fascinating properties of interacting quantum gases. 
When rotated at sufficiently large angular frequency, a superfluid 
develops vortex lines of zero density \cite{Feynman,Landau}, around which  
the circulation is quantized due to the single-valued character of
the corresponding wavefunction \cite{Onsager}. Quantized vortices constitute indeed 
one of the most important consequences 
of superfluidity, playing a fundamental role in various
physical systems, as superconductors \cite{Superconductors} and 
superfluid Helium \cite{Donelly}. Vortices and even vortex lattices have 
been created in alkali BECs in a series of milestone experiments 
\cite{Matthews99,Madison00,Abo-Shaeer01}. 

Similar to strings, vortex lines are 3D structures which 
may present transverse helical excitations, which 
are called Kelvin modes or waves \cite{Thompson1880}. These excitations 
were studied for quantized vortices in superfluids by 
Pitaevskii \cite{Pitaevskii1961}. Interestingly, the dispersion law 
for Kelvin modes at small wave vector $q$ follows a characteristic 
dependence $\epsilon(q) \sim -q^2\ln q\xi$, where $\xi$ is the healing
length. Kelvin modes play an important 
role in the physics of superfluid Helium \cite{Donelly,Ashton79}, 
and even of neutron stars  \cite{Epstein92}. Recently, Kelvin modes were 
experimentally observed in BEC \cite{Bretin2002}.


The DDI interaction may distort significantly the physics of vortices in
dipolar BECs, in particular the critical angular frequency 
for vortex creation \cite{ODell2005}. In addition, dipolar gases 
under fast rotation develop vortex lattices, which due to the DDI 
may be severely distorted \cite{Pu}, and
even may change its configuration from the usual triangular Abrikosov lattice
into other arrangements \cite{Cooper,Zhang2005}. 


In a recent Letter~\cite{Klawunn} we analyzed the case of a dipolar BEC in a 
one-dimensional optical lattice, and in particular the physics of straight vortex lines 
perpendicular to the two-dimensional planes defined by the lattice sites. We showed 
that due to the long-range character of the DDI, different parts of the vortex
line interact with each other, and hence the 3D character of the vortices
plays a much more important role in dipolar gases than in usual
short-range interacting ones. Specifically, we discussed 
that, interestingly, the DDI may severely modify the Kelvin-wave dispersion, which 
may even acquire a roton-like minimum. 
This minimum may touch zero energy for sufficiently 
large DDI and strong lattices, leading to the instability 
of the straight vortex line 
even for those situations in which the BEC as a whole is stable. However, the 
certainly very relevant question concerning the nature of this 
instability was not addressed in Ref.~\cite{Klawunn}.
In the present paper we discuss in detail this instability, 
showing that, interstingly,  it has a thermodynamical character, and it is linked to 
a second-order-like phase transition from a straight vortex line into an helical 
or snake-like vortex line depending on the dipole orientation.


The structure of the paper is as follows. In Sec.~\ref{sec:Model} we discuss
the effective model that describes a dipolar BEC in a sufficiently strong
optical lattice. In 
Sec.~\ref{sec:Kelvin-modes} we briefly summarize the main
results of~\cite{Klawunn}. In Sec.~\ref{sec:Helix} we discuss the phase
transition from straight into helical vortex-lines for the case of dipole
oriented parallel to the vortex line, whereas in Sec.~\ref{sec:Snake} we
discuss the transition into a snake-like vortex for dipoles perpendicular to
the vortex line. Finally, in Sec.~\ref{sec:Conclusions} we summarize our
conclusions.



\section{Dipolar BEC in an optical lattice. Effective model}
\label{sec:Model}

\begin{figure}[ht]
\begin{center}
\includegraphics[width=5.6cm,angle=270]{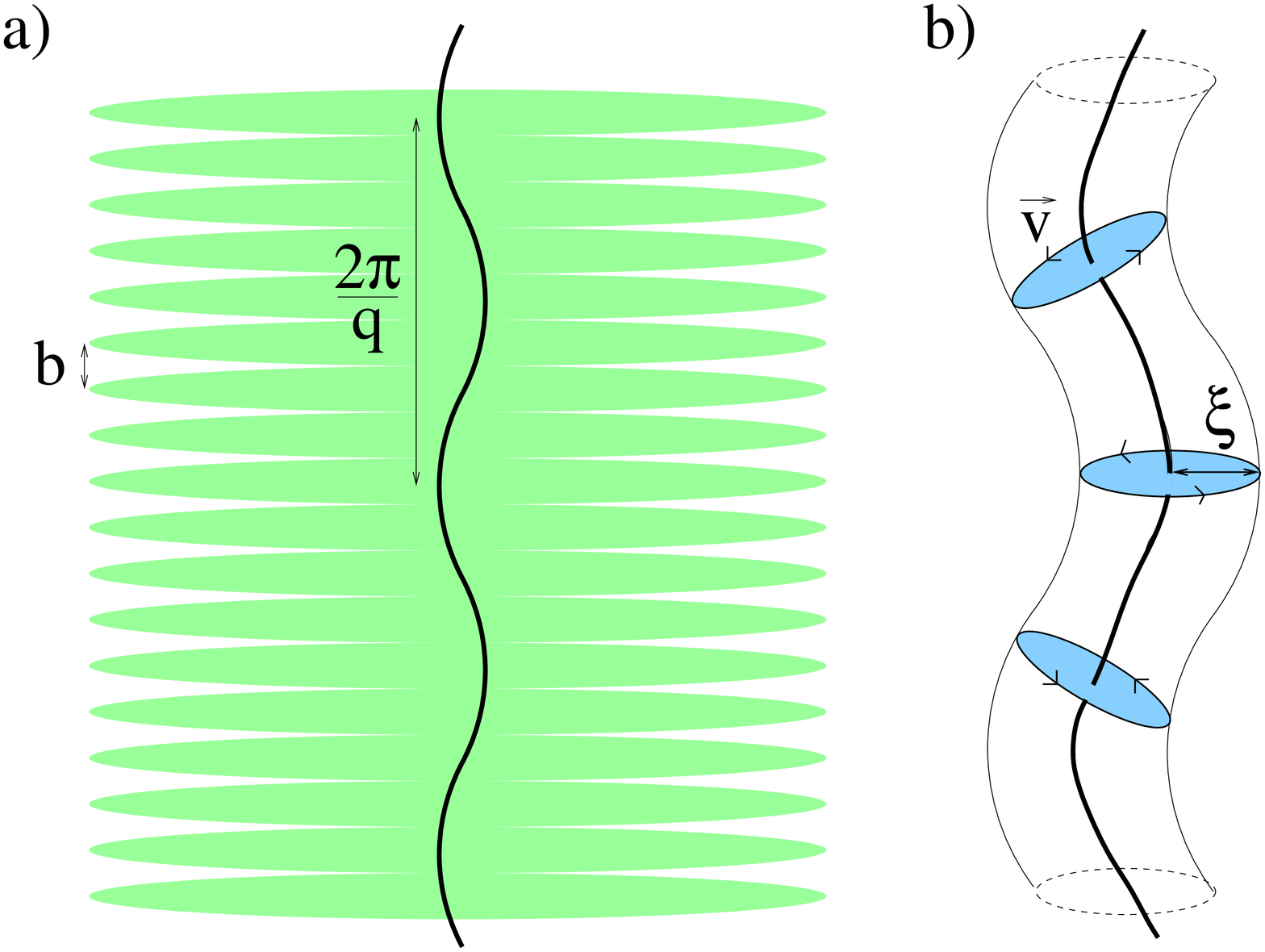}
\end{center}
\vspace*{-0.5cm}
\caption{(a) Vortex-line in an optical lattice with lattice spacing $b$.
The wave-number of the Kelvin mode is $q$,
whereas the corresponding wavelength satisfies $2 \pi /q \gg b$.
(b) The vortex-line has a finite core with healing length $\xi$
and a velocity field $\vec{v}$ perpendicular to the line.}
\vspace*{-0.2cm}
\label{fig:1}
\end{figure}

In the following, we consider a dipolar BEC of particles with mass $m$ and 
electric dipole $d$ (the results are equally valid for magnetic dipoles) oriented in the 
$z$-direction by a sufficiently large external field, and that hence 
interact via a dipole-dipole potential:
$V_d(\vec r)=\alpha d^2(1-3\cos^2\theta)/r^3$,
where $\theta$ is the angle formed by the vector 
joining the interacting particles and the dipole direction. The coefficient 
$\alpha$ can be tuned within the range $-1/2\leq\alpha\leq 1$ by rotating the external 
field that orients the dipoles much faster than any other
relevant time scale in the system~\cite{Tuning}. 
We consider the dipolar BEC as placed 
in a 1D optical lattice (see Fig.~\ref{fig:1}).  
At sufficiently low temperatures (and away from shape resonances \cite{Wang2006}) 
the physics of the dipolar BEC is provided by a non-local 
non-linear Schr\"odinger equation (NLSE) of the form:
\begin{equation}
\fl 
i\hbar \frac{\partial\Phi(\vec{r})}{\partial t}=
\left \{ -\frac{\hbar^2\nabla^2}{2m}  +  V_{\rm ol}(z)+ 
g|\Phi(\vec{r})|^2 
+ \int d \vec{r'} |\Phi(\vec{r'})|^2 V_d(\vec{r}-\vec{r}') \right \}\Phi(\vec{r}),
\label{GPgeneral}
\end{equation}
where $g=4\pi\hbar^2a/m$, with $a>0$ the $s$-wave scattering length, and 
$V_{\rm ol}(z)=sE_{R}\sin^2(Q_l z)$ is the 1D optical lattice, where 
$E_R=\hbar^2Q_l^2/2m$ is the recoil energy and $Q_l$ is the laser wave vector.

In the tight-binding regime (sufficiently strong lattice) 
we can write $\Phi(\vec{r},t)=\sum_jw(z-bj)\psi_j(\vec{\rho},t)$, 
where $b=\pi/Q_l$, $\vec{\rho}=\{x,y\}$ and $w(z)$ 
is the Wannier function associated to the 
lowest energy band. Substituting this Ansatz in Eq.~(\ref{GPgeneral}) we
obtain a discrete NLSE. We may then 
return to a continuous equation, where the presence of the lattice amounts for 
an effective mass $m^*$ along the lattice direction and for a renormalized
coupling constant $\tilde g$ \cite{Meret}: 
\begin{equation}
\fl i\hbar \frac{\partial\Psi(\vec{r})}{\partial t}=
\left \{ -\frac{\hbar^2\nabla_\perp^2}{2m}- \frac{\hbar^2\nabla_z^2}{2m^*} 
+ \tilde{g}|\Psi(\vec{r})|^2+
\int d \vec{r'} |\Psi(\vec{r'})|^2 V_d(\vec{r}-\vec{r}') 
\right \}\Psi(\vec{r}),
\label{GPeffe}
\end{equation}
where $\Psi(\vec r)=\psi_j(\vec\rho)/\sqrt{b}$ is the coarse-grained
wavefunction, $m^*=\hbar^2/2b^2J$ is the effective mass, with 
$J=\int w(z)[-(\hbar^2/2m)\partial^2_z+V_{\rm ol}(z)]w(z+b)dz$, 
and $\tilde{g}=bg\int w(z)^4dz+g_d\mathcal{C}$, with 
$g_d=\alpha 8\pi d^2/3$ and 
${\mathcal C}\simeq \sum_{j\ne 0} |\tilde w(2\pi j/b)|^2$, where
$\tilde w$ is the Fourier-transform of $w(z)$. 
The validity of Eq.~(\ref{GPeffe}) is limited to $z$-momenta 
$k_z\ll 2\pi/b$, 
in which one can ignore the discreteness of lattice. 
In addition, note that the single-band model breaks down
if the gap between the first and second band becomes comparable to 
other energy scales in the problem ($m/m^*\rightarrow 1$).

As mentioned above, the partially attractive character of
the DDI may lead to different types of instability in a dipolar BEC. 
We consider first an homogeneous 3D solution 
$\Psi_0(\vec r, t)=\sqrt{\bar n}\exp [-i\mu t/\hbar]$, where $\bar n$ denotes the 
condensate density, and $\mu=(g+\tilde V_d(0))\bar n$ the 
chemical potential, with $\tilde V_d(\vec k)=g_d[3k_z^2/|\vec k|^2-1]/2$.  
From the corresponding Bogoliubov-de Gennes (BdG) 
equations one gets that the energy $\epsilon(\vec k)$ corresponding 
to an excitation of wave number $\vec k$ fulfills: 
$
\epsilon(\vec k)^2=E_{\rm kin}(\vec k)[E_{\rm kin}(\vec k)+E_{\rm int}(\vec
k)]$, 
where $E_{\rm kin}(\vec k)=\hbar^2 k_\rho^2/2m+\hbar^2 k_z^2/2m^*$ is the kinetic energy, and 
$E_{int}(\vec k)=2(g+\tilde V_d(\vec k))\bar n$. Stable phonons (i.e. 
low-$k$ excitations) are only possible if 
$E_{int}>0$ for all directions, i.e. if  
$2+\beta (3k_z^2/|\vec k|^2-1)>0$, with  $\beta=g_d/\tilde{g}$. 
If $g_d>0$  phonons with $\vec k$ lying on the $xy$ plane are
unstable if $\beta>2$, while for $g_d<0$ 
phonons with $\vec k$ along $z$ are unstable if $\beta<-1$. 
Hence absolute phonon stability demands $-1<\beta<2$. 



\section{Kelvin-wave instability of a straight vortex line}
\label{sec:Kelvin-modes}

In this section we consider a straight vortex line along the 
$z$-direction (see Fig.~\ref{fig:1}). 
The corresponding wavefunction is of the form 
$\Psi_0(\vec r,t)=\phi_0(\rho)\exp(i\varphi)\exp [-i\mu t/\hbar]$, where $\varphi$ is
the polar angle on the $xy$ plane. The function $\phi_0(\rho)$ fulfills 
\begin{equation}
\mu\phi_0(\rho)=\frac{\hbar^2}{2m}\left(-\frac{1}{\rho}\partial_\rho\rho\partial_\rho
+\frac{1}{\rho^2}\right)
\phi_0(\rho)+\bar{g}|\phi_0(\rho)|^2\phi_0(\rho),
\label{phi0}
\end{equation}
where $\bar{g}=\tilde{g}-g_d/2$.
Note that, due to the homogeneity of $\Psi_0$ 
in the $z$-direction, the DDI just regularizes the local term. 
The density of the vortex core is given by $|\phi_0(\rho)|^2$, which vanishes
at $\rho=0$ and becomes equal to the bulk
density $\bar{n}$ at distances larger than the healing length 
$\xi=\hbar/\sqrt{m\bar{g} \bar{n}}$ (see Fig.~\ref{fig:1}). Note that $\xi$ 
depends on the DDI. 

In the following we consider Kelvin-wave excitations of the straight vortex
line of the form (see Fig.~\ref{fig:1})
$\Psi(\vec r,t)=\Psi_0(\vec r,t)+\chi(\vec r,t)\exp [i (\varphi-\mu t/\hbar)]$, 
where \cite{Pitaevskii1961} 
$\chi(\vec r,t)=\sum_{l,q} [ u_l (\rho) \exp [i(\varphi l+qz-\epsilon t/\hbar)]-v_l(\rho)^*
\exp [-i(\varphi l+qz-\epsilon^* t/\hbar)]]$ (as mentioned in
Sec.~\ref{sec:Model}, $q\ll 2\pi/b$ in order to justify the validity of the
effective model). 
introducing this Ansatz into~(\ref{GPeffe}) and linearizing in $\chi$, one obtains
the corresponding BdG equations:
\begin{eqnarray}
\fl\epsilon u_l(\rho)= \left[ \frac{\hbar^2}{2m}
\left(-\frac{1}{\rho}\partial_\rho\rho\partial_\rho
 + \frac{(l+1)^2}{\rho^2}+\frac{m}{m^*}q^2\right) -\mu+ 2\bar g \psi_0(\rho)^2 \right]u_l(\rho) - \bar g \psi_0(\rho)^2 v_l(\rho) \nonumber \\
 +\frac{3\beta}{2-\beta}\bar g  q^2\int_0^\infty d\rho' \rho' \psi_0(\rho') \psi_0(\rho)
\left[u_l(\rho')-v_l(\rho') \right]F_l(q\rho,q\rho') \label{BdG1} \\
\fl\epsilon v_l(\rho)=
 -\left[ \frac{\hbar^2}{2m}\left(-\frac{1}{\rho}\partial_\rho\rho\partial_\rho
 + \frac{(l-1)^2}{\rho^2}+\frac{m}{m^*}q^2\right) -\mu+ 2\bar g \psi_0(\rho)^2 \right]v_l(\rho)
+ \bar g \psi_0(\rho)^2 u_l(\rho) \nonumber \\
 +\frac{3\beta}{2-\beta}\bar g  q^2\int_0^\infty d\rho' \rho' \psi_0(\rho') \psi_0(\rho)
\left[u_l(\rho')-v_l(\rho') \right]F_l(q\rho,q\rho'), \label{BdG2}
\end{eqnarray}
with $F_l(x,x')=I_l(x_<)K_l(x_>)$ ,
where $I_l$ and $K_l$ are modified Bessel functions, and 
$x_>=$max$(x,x')$, $x_<=$min$(x,x')$. For every $q$ we determine the lowest 
eigen-energy $\epsilon$, that provides the dispersion law discussed below.

\begin{figure}
\begin{center}
\includegraphics[width=6.8cm,angle=270]{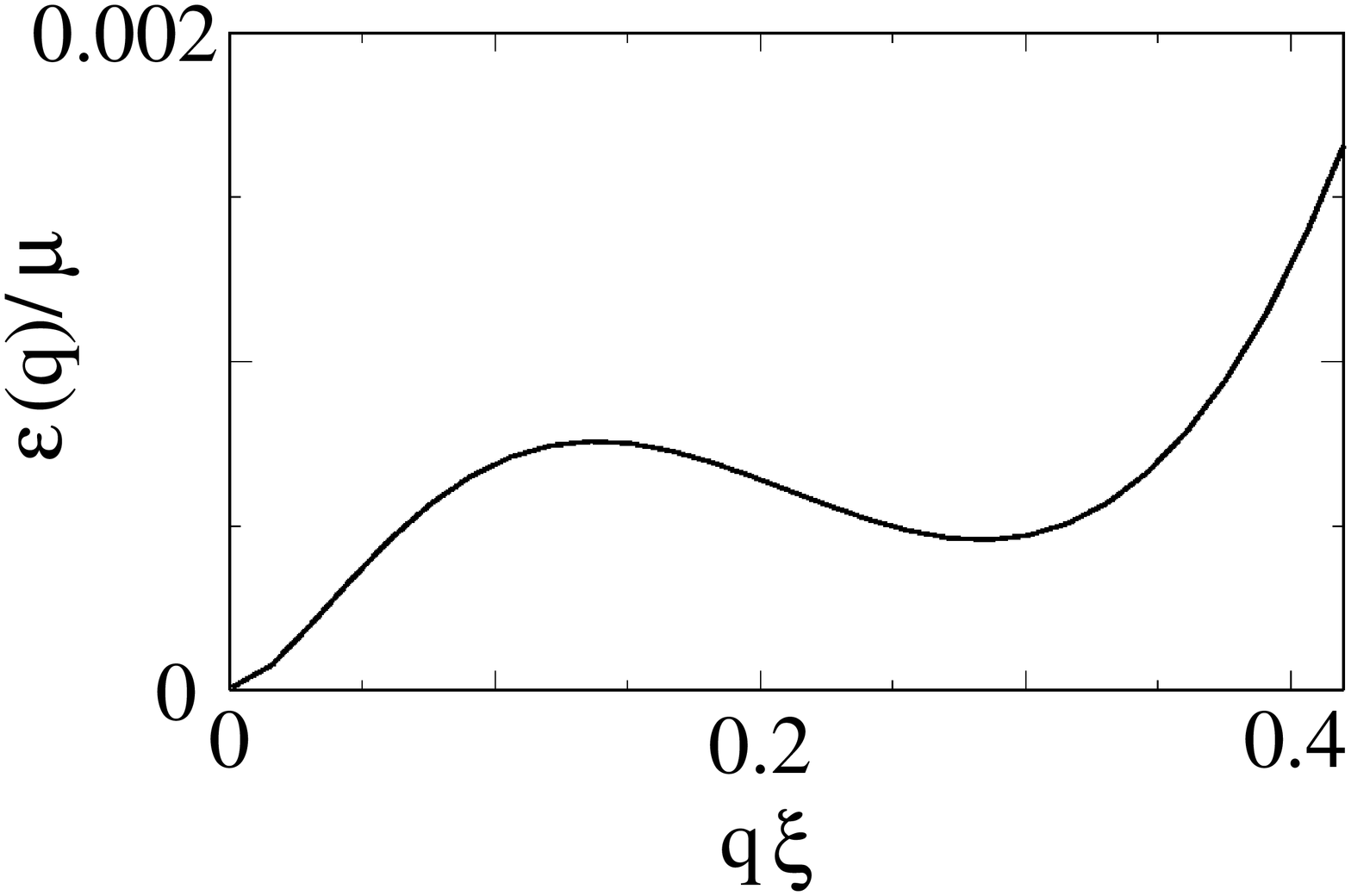}
\end{center}
\vspace*{-0.5cm}
\caption{Dispersion $\epsilon(q)$ as a function of $q\xi$ for $m/m^*=0.143$,
and $\beta=-0.8$.}
\vspace*{-0.2cm}
\label{fig:2}
\end{figure}


The first line at the rhs of Eqs. (\ref{BdG1}) and (\ref{BdG2}) is exactly the
same as that expected for a vortex in a short-range interacting BEC
\cite{Pitaevskii1961}, but with
the regularized value $\bar g$. Hence,  
in absence of DDI (or equivalently from Eqs.~(\ref{BdG1})
and (\ref{BdG2}) without the last integral term) the dispersion law
at low momenta ($q\xi\ll 1$) is provided by the well-known expression
$\epsilon(q)=-(\hbar^2q^2/2m^*)\ln \left[ (m/m^*)^{1/2}q\xi\right]$.
On the contrary, the last term at the rhs of both equations
is directly linked to the long-range character of the DDI and, as we show
below, leads to novel phenomena in the physics of Kelvin modes ($l=-1$) in dipolar
BECs. In the following, and in order to isolate the effect of the DDI on
the core size with respect to the effect of the integral terms in Eqs. 
(\ref{BdG1}) and (\ref{BdG2}), we fix $\bar g$ and change
the parameter $\beta$ which is proportional to the dipole-dipole coupling constant. 

\begin{figure}
\begin{center}
\includegraphics[width=8.3cm,angle=270]{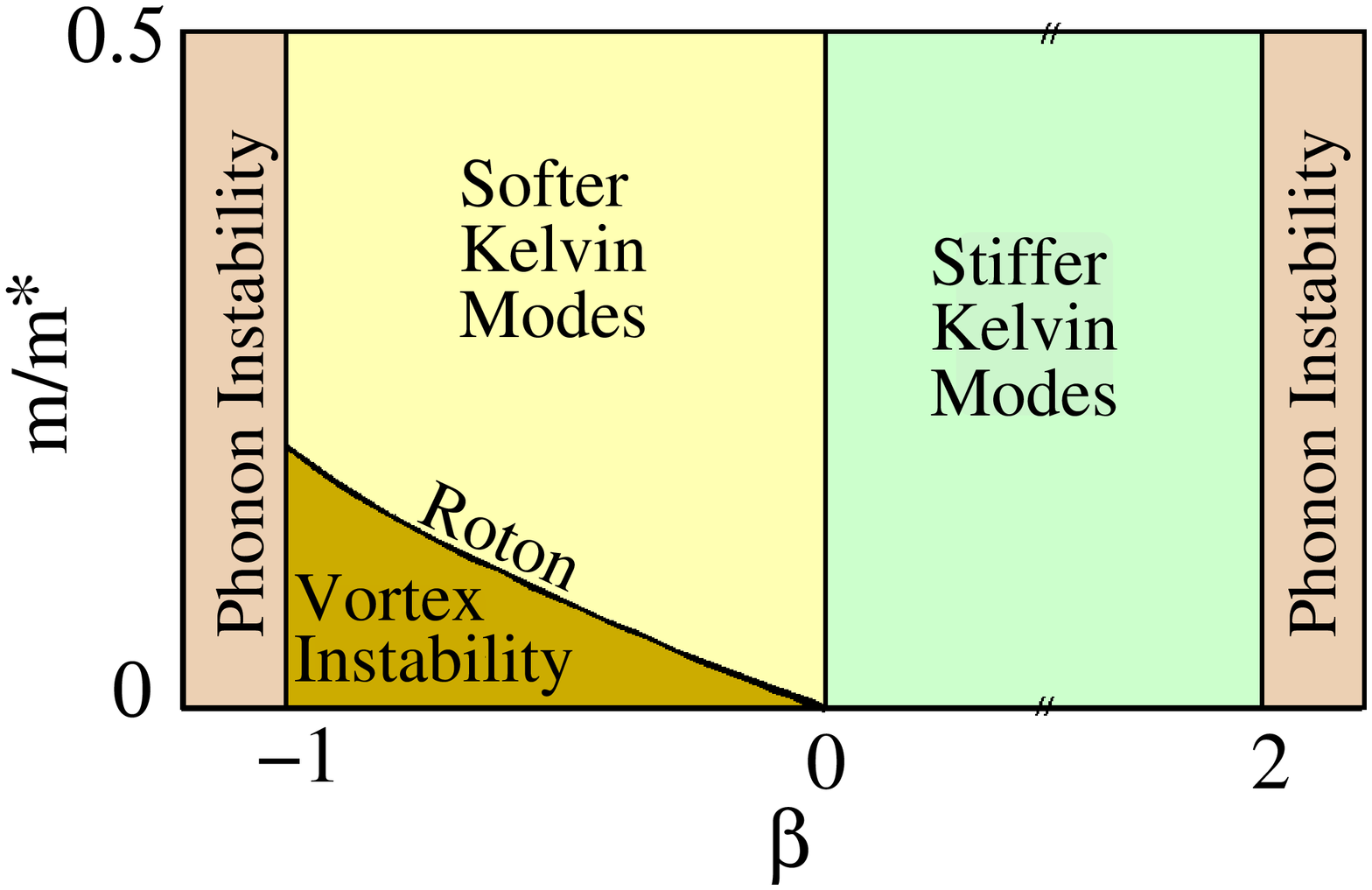}
\end{center}
\vspace*{-0.6cm}
\caption{Stable/unstable regimes for straight vortex lines.}
\vspace*{-0.3cm}
\label{fig:3}
\end{figure}

The integral terms of Eqs. (\ref{BdG1}) and (\ref{BdG2}) significantly
modify the Kelvin-mode spectrum in a different way depending whether $\beta>0$
or $\beta<0$. The different regimes are summarized in Fig.~\ref{fig:3}. 
As shown in Ref.~\cite{Klawunn}, for $\beta>0$ the excitation energy increases, 
i.e. the vortex line becomes stiffer against transverse modulations. 
The opposite occurs for $\beta<0$, i.e. the Kelvin-modes become softer.
However, the softening of the Kelvin-wave spectrum never leads to instability 
in absence of an additional lattice ($m=m^*$), since destabilization 
demands $\beta<-1$, for which, as we show in Sec.~\ref{sec:Model} 
the whole dipolar BEC is phonon-unstable. Interestingly, an increase 
of the potential depth of the additional lattice leads to a reduction of  
the role of the kinetic energy term $mq^2/m^*$ in Eqs. (\ref{BdG1}) 
and (\ref{BdG2}) that enhances the effect of the dipolar interaction on the
dispersion law. As a consequence, as shown in Fig.~\ref{fig:2}, 
in addition to the $-q^2\ln (q\xi)$ dependence at low $q$, 
a roton-like minimum eventually appears at intermediate $q$. 
For a sufficiently small $(m/m^*)_{cr}$ (thick curve in Fig.~\ref{fig:3}) 
the roton minimum reaches zero energy. For $m/m^*\leq (m/m^*)_{cr}$ 
the energy at the roton becomes negative, and hence  
the straight vortex-line becomes thermodynamically unstable against 
the formation of a new vortex-line configuration, 
which we discuss in the following sections. 

The intriguing dependence of the dispersion of Kelvin-waves on $m/m^*$ 
and $\beta$ is explained by the competition of the two  
processes involved in the inter-site physics, namely tunneling and intersite DDI. 
On one side, the hopping energy
is minimized when vortex cores at neighboring sites are placed right 
on top of each other. 
Hence tunneling tends to maintain the vortex line straight. On the other side, 
we may obtain an intuitive picture concerning the intersite DDI  
by sketching the vortex core as a 1D chain of dipolar holes. 
Dipolar holes interact in exactly the same way as dipolar particles, 
and hence for $\beta>0$ attract each other maximally when 
aligned along the dipole direction, i.e. the $z$-direction. As a consequence, 
for $\beta>0$ DDI and hopping add up in keeping the vortex straight, and  
the vortex line becomes stiffer. On the contrary, for $\beta<0$ 
the cores maximally repel each other when aligned along the $z$-direction. Hence
DDI and tunneling compete, and the vortex line becomes softer. 

\section{Helical vortices}
\label{sec:Helix}

From the intuitive picture discussed in the previous section it becomes 
clear that when $m/m^*$ becomes sufficiently small the tunneling cannot 
balance the DDI any longer, and as a consequence the straight vortex 
line becomes thermodynamically unstable, as already mentioned in 
our discussion of the Bogoliubov analysis in the previous section.
In this section we show that
this instability is related to a transition from a straight vortex into a 
twisted vortex-line. This twisting, which we confirm below numerically, may 
be expected by analyzing the DDI between vortex cores at neighboring sites. 
This energy is minimized by laterally displacing the vortices 
with respect to each other a finite distance on the $xy$-plane. These lateral 
displacements between nearest neighbors lead to a twisting of the vortex line. 
Note that due to the symmetry of the problem, we may expect an helical structure, 
as we confirm in the following.

\begin{figure}
\begin{center}
\includegraphics[width=7.8cm,angle=270]{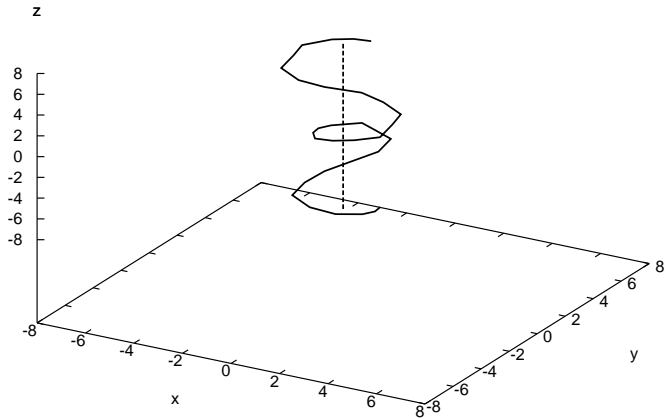}
\end{center}
\vspace*{-0.5cm}
\caption{Vortex-line configurations of the ground state for $\beta=-0.8$
and $m/m^*=0.15$ (straight) and $m/m^*=0.04$ (helical), respectively.
For both cases the dipole is assumed in the $z$-direction.
The unit of length is $\sqrt{2}\xi$}
\vspace*{-0.2cm}
\label{fig:4}
\end{figure}

In order to determine the new lowest-energy configuration of the vortex line, 
we have performed fully 3D numerical calculations of the GPE (\ref{GPeffe}). We start our
calculations from an initially imposed straight vortex line and let the system
evolve in imaginary time. We performed our calculations employing a
cylindrical numerical box placed several healing lengths apart from the vortex 
core to minimize boundary effects. The cylindrical box configuration permits
a sufficiently flat density in the region where the vortex core is created,
allowing us to avoid unwanted density effect, which will obscure the analysis of the
vortex stability.
As expected, the imaginary time evolution leads to  
a ground-state straight vortex line configuration for those values of 
$\beta$ and $m/m^*$ lying inside the stable regime of Fig.~\ref{fig:3}. 
On the contrary, for those regions within the instability regime a
qualitatively new ground-state is found, where we observe a departure 
from the straight form into an helical configuration, as in Fig.~\ref{fig:4}.

\begin{figure}
\begin{center}
\includegraphics[width=7.8cm,angle=0]{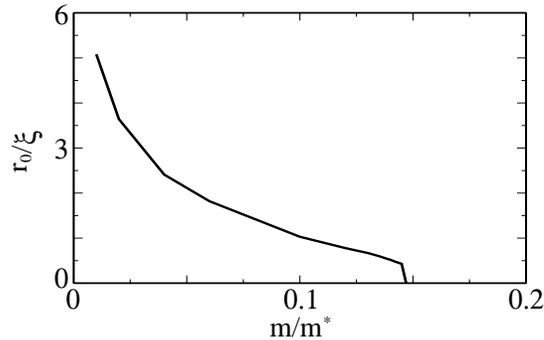}
\end{center}
\vspace*{-0.5cm}
\caption{Helix amplitude $r_0$ (see text) 
as a function of $m/m^*$ for $\beta=-0.8.$}.
\vspace*{-0.2cm}
\label{fig:5}
\end{figure}

Hence at the line $(m/m^*)_{cr}$ there is a phase transition from straight
into helical ground-state vortex lines. In order to characterize this phase 
transition we have performed an exhaustive analysis of the helical
configurations, which may be characterized by a pitch with wavenumber $Q$ and 
a radius $r_0$. We confirmed that the wave-number $Q$ is indeed 
comparable to the minimum
$q_{\rm rot}$ in the dispersion law obtained from the Bogoliubov analysis of 
Sec.~\ref{sec:Kelvin-modes}. For several fixed values of $\beta$ 
we performed a large number of 3D numerical simulations for different $m/m^*$, to 
analyze the behavior of the amplitude $r_0$ inside the unstable region 
and at the transition to the stable region. Our results 
(for $\beta=-0.8$) are depicted in
Fig.~\ref{fig:5}. We observe a gradual decrease of $r_0$ when $m/m^*$
approaches the stability border $(m/m^*)_{cr}$. For values to close 
to the stability border $r_0\ll\xi$ and hence it becomes in practive
(both numerically and experimentally) impossible to
discern a significant vortex twisting. Our simulations are hence compatible 
with a second-order-like phase-transition from an helical into a straight 
lowest-energy vortex-line configuration.

Note that inside the instability region an increase of the lattice strength
(i.e. a decrease of $m/m^*$) results in a larger bending which comes together with a
shallower binding energy for the line. If the binding energy becomes lower 
than other typical energy scales involved in the system (inhomogeneity, boundary 
effects) then the 3D vortex line breaks into uncorrelated 2D vortices. 
The latter is, of course, also expected in the absence of DDI if the tunneling 
becomes sufficiently small. Note that in Fig.~\ref{fig:3} we did not consider 
any other additional energy scale, and hence the 3D vortex line could be considered  
all the way down to very small $m/m^*$. However, in our numerical simulations 
we do have boundary effects introduced by our finite cylindrical numerical box. 
As a consequence we have observed (also for the results of next section) 
that for very small $m/m^*$ inside the instability
regime the boundary effects may completely unbound the vortex line into
uncorrelated 2D vortices at each layer. However, in the present paper we are more
concerned with the phase transition boundary, where the binding is still
dominant even when the vortex is bent.

\section{Snake-like vortices}
\label{sec:Snake}

In Secs.~\ref{sec:Kelvin-modes} and \ref{sec:Helix}, 
we have discussed the case in which the dipoles are 
oriented along the $z$-direction. In that situation, the
cylindrical symmetry of the problem largely simplified the Bogoliubov analysis. 
The helical instability demands $\beta<0$,
and it is hence necessary the employ of 
the tuning mechanism discussed in Sec.~\ref{sec:Model}. 
However, the tuning is not strictly necessary to observe
the destabilization of the straight vortex line. 
A perhaps experimentally simpler scenario 
is offered by the case in which the dipoles are oriented 
perpendicular to the vortex line (and hence also perpendicular to the
over-imposed lattice potential). 
Then the DDI is again repulsive in $z$-direction
and competes with the kinetic energy in what concerns the 
stability of the straight vortex-line 
in a similar way as described in the previous sections.
In this section we analyze this particular
case (orientation along $y$). We show that the vortex line may be also
destabilized in this experimentally less involved case.

As for Sec.~\ref{sec:Helix}, we have evolved 
an initially straight vortex in imaginary time using
Eq.~(\ref{GPeffe}) employing similar numerical conditions. 
However, the analysis in the new configuration is  
largely handicapped due to the distortion of the density of the condensate 
induced by the fact that the dipole in now on the $xy$ plane, and hence breaks the 
polar symmetry. These density modulations may be reduced (but not
fully eliminated) by considering larger numerical boxes. However, the latter 
may make the 3D numerical simulation prohibitively long. 

\begin{figure}
\begin{center}
\includegraphics[width=7.8cm,angle=270]{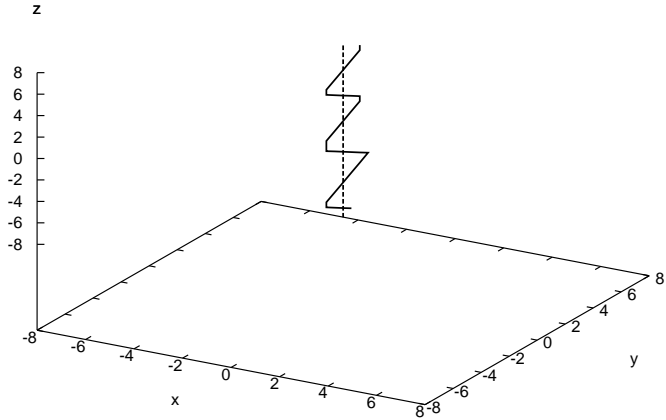}
\end{center}
\vspace*{-0.5cm}
\caption{Vortex-line configurations of the ground state for $\beta=1.2$
and $m/m^*=0.075$ (straight) and $m/m^*=0.04$ (snake-like), respectively.
For both cases the dipole is assumed in the $y$-direction.
The unit of length is $\sqrt{2}\xi$}
\vspace*{-0.2cm}
\label{fig:6}
\end{figure}

In spite of these difficulties we observed the departure from the straight 
vortex line for sufficiently large $\beta$ 
(note that in this configuration 
$\beta>0$ and hence it may take values up to $2$ before entering into the
regime of phonon instability discussed in Sec.~\ref{sec:Model}).
Due to the broken cylindrical symmetry, 
for sufficiently large $\beta$ and small $m/m^*$ the straight vortex line is destabilized 
into a snake-like configuration on the $yz$ plane rather than into an helix,
as in the previous section, as shown in Fig.\ref{fig:6}. We have analyzed 
the amplitude of the helix when approaching the stability regime. Our results
are compatible with a second-order-like behavior as in the previous section, 
although, as mentioned above, a rigorous analysis is largely handicapped 
by the appearance of strong density modulations in the simulations.

\section{Conclusions}
\label{sec:Conclusions}

In this paper we have studied the physics of vortex lines in dipolar gases. In
particular, we have analyzed in detail the stability of a straight vortex 
line with respect to Kelvin waves. We have shown that the presence of 
an additional optical lattice along the vortex line may allow for the 
observation of a dipole-induced destabilization of the straight vortex line
due to the softening of a roton minimum in the Kelvin-wave spectrum. If this
occurs the straight vortex-line configuration ceases to be that of minimal
energy, and there is a second-order-like phase transition into an helical 
or snake vortex-line, depending on the dipole orientation.

\ack
We thank P. Pedri and R. Nath for useful discussions.
This work was supported by the DFG (SFB407, SPP1116), and the ESF
(EUROQUASAR). 


\end{document}